\documentclass[amsmath,amssymb,12pt,aps]{revtex4-2}
\usepackage{amsfonts}
\usepackage[utf8]{inputenc} 
\usepackage{graphicx}
\usepackage{float}
\usepackage{natbib}
\usepackage{wrapfig}
\usepackage{xcolor}
\usepackage{dcolumn}
\usepackage[english]{babel}
\begin{document}
	
\title{Third order interactions shift the critical coupling in multidimensional Kuramoto models}

\author{Ricardo Fariello}

\affiliation{Departamento de Ci\^encias da Computa\c{c}\~ao,
	Universidade Estadual de Montes Claros, 39401-089, Montes Claros, MG, Brazil} 
	
\author{Marcus A.~M.~de Aguiar}

\affiliation{Instituto de F\'isica Gleb Wataghin, Universidade Estadual de Campinas, Unicamp 13083-970, Campinas, SP, Brazil}

\begin{abstract}
	
The study of higher order interactions in the dynamics of Kuramoto oscillators has been a topic of intense recent research. Arguments based on dimensional reduction using the Ott-Antonsen ansatz show that such interactions usually facilitate synchronization, giving rise to bi-stability and hysteresis. Here we show that three body interactions shift the critical coupling for synchronization towards higher values in all dimensions, except $D=2$, where a cancellation occurs. After the transition, three and four body interactions combine to facilitate synchronization. Similar to the 2-dimensional case, bi-stability and hysteresis develop for large enough higher order interactions. We show simulations in $D=3$ and $4$ to illustrate the dynamics.
	
\end{abstract}

\maketitle
	
\section{Introduction}

Pairwise interactions are not enough to describe the dynamics of many complex systems \cite{battiston2020networks}. In several cases actions take place at the level of groups of agents, as found in neuroscience \cite{petri2014homological,sizemore2018cliques,ganmor2011sparse}, 
ecology \cite{grilli2017higher}, biology \cite{sanchez2019high} and social sciences \cite{benson2016higher,de2020social}. Efforts to account for interactions beyond the usual two body terms have generated a large body of literature in several areas, particularly in propagation of epidemics \cite{iacopini2019simplicial,jhun2019simplicial,vega2004fitness} and synchronization \cite{berec2016chimera,skardal2019abrupt,skardal2020higher,sarika2024,biswas2024symmetry,sayeed2024global,muolo2024phase}.

In the specific case of the Kuramoto model, it has been shown that higher order terms create bi-stable regions in parameter space, and therefore hysteresis, a feature that does not exist in the original model with pair interactions \cite{skardal2020higher}. For fully connected networks and Lorentzian distribution of natural frequencies, where the Ott-Antonsen ansatz \cite{Ott2008} can be applied, 
theory predicts that the critical coupling for synchronization does not change, but a line of saddle-node bifurcation marks the appearance of a new synchronized equilibrium that co-exist with the asynchronous state. Moreover,
above the critical coupling, higher order terms facilitate synchronization.

Here we consider the multidimensional version of the Kuramoto model, as proposed in \cite{2019continuous,lipton2021kuramoto}. We first derive the higher order corrections in $D$ dimensions that generalize the corresponding interactions in $D=2$. Then, we show that, generally, the critical coupling for synchronization changes when three-body interactions are taken into account, making the transition 
from disordered to ordered states more difficult. Only in two dimensions, corresponding to the original Kuramoto model, this dependence disappears. Above the modified critical coupling third and fourth order terms combine to facilitate synchronization. These features are demonstrated using mean-field arguments and are verified with numerical simulations, since the Ott-Antonsen ansatz 
does not extend to dimensions larger than $2$.

\section{Higher order interactions in the multidimensional Kuramoto model}

In the Kuramoto model, a set of $N$ oscillators, represented only by their phases $\theta_i$, are coupled according to the equations
\begin{eqnarray}
	\dot{\theta}_i &=& \omega_i + \frac{k_1}{N} \sum_{j=1}^N \sin{(\theta_j-\theta_i)} 
	\label{kuramotoori}
\end{eqnarray}
where $\omega_i$ are their natural frequencies, selected from a symmetric distribution $g(\omega)$. Three and four body interactions are introduced as
\begin{eqnarray}
	\dot{\theta}_i &=& \omega_i + \frac{k_1}{N} \sum_{j=1}^N \sin{(\theta_j-\theta_i)} + 
		\frac{k_2}{N^2} \sum_{j,k=1}^N \sin{(2\theta_j-\theta_k -\theta_i)} \nonumber \\
		&+& \frac{k_3}{N^3} \sum_{j,k,m=1}^N \sin{(\theta_j+\theta_k -\theta_m-\theta_i)}
	\label{kuramoto}
\end{eqnarray}
where $i = 1, ..., N$ and $k_1$, $k_2$ and $k_3$ are the coupling constants for pairwise (1-simplex), triplets (2-simplex) and quadruplets (3-simplex) interactions. 
We note that the specific form of the three and four body interactions in 
Eq.~(\ref{kuramoto}) follows the choice in \cite{skardal2020higher}, although other, more symmetric forms, have also been considered \cite{battiston2020networks,ashwin2016hopf}.

Using the definition of the order parameter
\begin{equation}
	z = \frac{1}{N} \sum_j e^{i \theta_j} = r e^{i \psi}
	\label{zdef}
\end{equation}
we obtain
\begin{equation}
\frac{1}{N} \sum_j e^{i (\theta_j - \theta_i)} = r e^{i (\psi - \theta_i)},
\end{equation}
\begin{equation}
	\frac{1}{N} \sum_k e^{i (\theta_k + \theta_i - 2\theta_j)} = r e^{i (\psi + \theta_i - 2\theta_j)}
\end{equation}
and
\begin{equation}
	z^2 z^* e^{-i\theta_i}= \frac{1}{N^3} \sum_{j,k,m} e^{i (\theta_j +\theta_k -\theta_m- \theta_i)} = r^3 e^{i (\psi - \theta_i)}.
\end{equation}

Taking the imaginary part of these equations we can rewrite the Kuramoto model as
\begin{equation}
	\dot{\theta}_i = \omega_i + k_1 r \sin(\psi - \theta_i) +  k_2 r \frac{1}{N} \sum_{j=1}^N \sin{(2\theta_j-\theta_i-\psi)} + k_3 r^3 \sin(\psi-\theta_i). 
	\label{kuramoto23}
\end{equation}

Following \cite{skardal2020higher} we also define a second order parameter
\begin{equation}
	z_2 = \frac{1}{N} \sum_j e^{2 i \theta_j} = p_2 e^{i \xi_2}
	\label{z2def}
\end{equation}
and the related vector
\begin{equation}
	\vec{p}_2 = p_2 (\cos\xi_2,\sin\xi_2)
\end{equation}
so that Eq.~(\ref{kuramoto23}) becomes
\begin{equation}
	\dot{\theta}_i = \omega_i + k_1 r \sin(\psi - \theta_i) +  k_2 r p_2 \sin{(\xi_2-\theta_i-\psi)} + k_3 r^3 \sin(\psi-\theta_i). 
	\label{kuramoto24}
\end{equation}
We use $p_2$ and $\xi_2$ for the module and phase of $z_2$ as the vector $\vec{r}_2$ will be reserved for a later definition.

Equation~(\ref{kuramoto24}) can be generalized to more dimensions if we can write it in terms of the unit vectors $\vec{\sigma_i} = (\cos{\theta_i},\sin{\theta_i}) \equiv (\sigma_{ix},\sigma_{iy})$ \cite{2019continuous,barioni2021complexity}. Computing $\dot{\sigma}_{ix} = - \dot{\theta}_i \sin \theta_i$ and using Eq.~(\ref{kuramoto24}) we find
\begin{eqnarray}
	\dot{\sigma}_{ix} &=& - \sin \theta_i[ \omega_i + k_1 r \sin(\psi - \theta_i) +  k_2 r p_2 \sin{(\xi_2-\theta_i-\psi)} + k_3 r^3 \sin(\psi-\theta_i)]  \nonumber \\
	&\equiv & -\omega_i \sigma_{iy} + k_1 T_{1x} + k_2 T_{2x} + k_3 T_{3x}.
\end{eqnarray}
The contributions of the interaction terms are computed as follows:
\begin{eqnarray}
	T_{1x} & = & - r \sin \theta_i \sin(\psi - \theta_i) \nonumber \\
	&=& - r  \sin\psi \sin \theta_i \cos\theta_i + r \sin\theta_i^2 \cos\psi \nonumber \\
	&=&  r \cos\psi - r [ \sin\psi \sin \theta_i  +  \cos\theta_i \cos\psi] \cos\theta_i \nonumber \\
	&=& r_{x} -(\vec{r} \cdot \vec{\sigma}_i) \sigma_{ix}
\end{eqnarray}
where we defined the vector
\begin{equation}
	\vec{r} = (r\cos\psi,r\sin\psi)
\end{equation}
representing the center of mass of the system.

The third order term is $T_{2x} =  - r p_2 \sin \theta_i \sin(\xi_2-\psi - \theta_i) $, which becomes identical to $T_{1x}$ if we replace $r \rightarrow r p_2$ and $\psi \rightarrow \xi_2 - \psi$. Therefore we find
\begin{eqnarray}
	T_{2x} 	&=& r_{2x} -(\vec{r}_2 \cdot \vec{\sigma}_i) \sigma_{ix}
\end{eqnarray}
where
\begin{equation}
	\vec{r}_2 = r p_2(\cos(\xi_2-\psi), \sin(\xi_2-\psi)) = (r_x p_{2x}+r_y p_{2y}, r_x p_{2y}-r_y p_{2x}).
\end{equation}
Similarly we obtain
\begin{eqnarray}
	T_{3x} 	&=& r_{3x} -(\vec{r}_3 \cdot \vec{\sigma}_i) \sigma_{ix}
\end{eqnarray}
where
\begin{equation}
	\vec{r}_3 = r^3 (\cos\psi, \sin\psi) = r^2 \vec{r}
\end{equation}
with $r_3 = r^3$ and $\psi_3 = \psi$. 

After doing a similar calculation for the $\sigma_{iy}$ we can write the dynamical equations in vector form as 
\begin{equation}
	\frac{d \vec{\sigma_i}}{d t} = \mathbf{W}_i \vec{\sigma_i} + \sum_{j=1}^3 k_j [\vec{r}_j -(\vec{\sigma}_i \cdot \vec{r}_j) \vec{\sigma}_i]
	\label{eq3}
\end{equation}
where we identify $\vec{r}_1 \equiv \vec{r}$ and  $\mathbf{W}_i$ is the anti-symmetric matrix 
\begin{equation}
	\mathbf{W}_i = \left( 
	\begin{array}{cc}
		0 & -\omega_i \\
		\omega_i & 0
	\end{array}
	\right).
	\label{wmat}
\end{equation}

In order to extend the model to higher dimensions it is essential that all vectors $\vec{r}_j$ can be written in terms of the unit vectors $\vec{\sigma}_j$.
Clearly
\begin{equation}
	\vec{r} = \frac{1}{N} \sum_{k=1}^N \vec{\sigma}_k
\end{equation}
and
\begin{equation}
	\vec{r}_3 = r^2 \vec{r}.
\end{equation}
To write $\vec{r}_2$ in terms of the $\vec{\sigma}_j$'s we first realize that
\begin{equation}
	p_{2x} = \frac{1}{N} \sum_j \cos 2\theta_j = \frac{1}{N} \sum_j [\sigma_{jx}^2 - \sigma_{jy}^2] = \frac{2}{N} \sum_j \sigma_{jx}^2 - 1 
	    \equiv  2\langle \sigma_x^2 \rangle - 1  = 1 - 2 \langle \sigma_y^2 \rangle 
	    \label{p2x}
\end{equation}
and
\begin{equation}
	p_{2y} = \frac{1}{N} \sum_j \sin 2\theta_j = \frac{2}{N} \sum_j  \sigma_{jx} \sigma_{jy} \equiv  2\langle \sigma_x \sigma_y \rangle .
	\label{p2y}
\end{equation}
Next, for any vector $\vec{\sigma}_j$ we define the dyadic matrix $\Lambda_j = \vec{\sigma}_j \vec{r}^T$ and the vector
\begin{equation}
	\frac{1}{N} \sum_{j} \Lambda_{j} \vec{\sigma}_j = 
   \frac{1}{N} \sum_{j} \left(
	\begin{array}{cc}
		\sigma_{jx} r_{x} & \sigma_{jx} r_{y} \\
		\sigma_{jy} r_{x} & \sigma_{jy} r_{y} \\		 
	\end{array}	
	\right)
	\left(
	\begin{array}{c}
		\sigma_{jx} \\
		\sigma_{jy}
	\end{array}
	\right) =
	\frac{1}{N} \sum_{j} \left(
	\begin{array}{c}
		r_{x} \sigma_{jx}^2 + r_{y} \sigma_{jx} \sigma_{jy} \\
		r_{x} \sigma_{jx} \sigma_{jy} + r_{y} \sigma_{jy}^2
	\end{array}
	\right)	
\end{equation}
which can also be written as
\begin{equation}
	 \left(
	\begin{array}{c}
		r_{x} \langle \sigma_{x}^2  \rangle + r_{y}  \langle \sigma_{x} \sigma_{y} \rangle \\
		r_{x} \langle \sigma_{x} \sigma_{y} \rangle + r_{y} \langle \sigma_{y}^2  \rangle
	\end{array}
	\right)	= \frac{1}{2} \left(
	\begin{array}{c}
		r_{x} (1+p_{2x}) + r_{y}  p_{2y} \\
		r_{x} p_{2y} + r_{y} (1-p_{2x}) 
	\end{array}
	\right)	 =\frac{1}{2} (\vec{r} + \vec{r}_2)
\end{equation}
where we used Eqs.~(\ref{p2x}) and (\ref{p2y}). Therefore
\begin{equation}
	\frac{1}{N} \sum_{j} \Lambda_{j} \vec{\sigma}_j  
	= \frac{1}{N} \sum_j (\vec{r} \cdot \vec{\sigma}_j) \vec{\sigma}_j = \frac{1}{2}(\vec{r}+\vec{r}_2)
\end{equation}	
or
\begin{equation}
\vec{r}_2 =  \frac{2}{N} \sum_j (\vec{r} \cdot \vec{\sigma}_j) \vec{\sigma}_j  - \vec{r}.
\end{equation}	

Equation~(\ref{eq3}) can now be extended to higher dimensions by simply considering unit vectors $\vec\sigma_i$ in $D$-dimensions, rotating on the surface of the corresponding $(D-1)$ unit sphere \cite{2019continuous}. Particles are now represented by $D-1$ spherical angles, generalizing the single phase $\theta_i$ of the original model. The matrices $\mathbf{W}_i$ become $D \times D$ anti-symmetric matrices containing the $D(D-1)/2$ natural frequencies of each oscillator. 
The equations up to fourth order can be summarized as follows:
\begin{equation}
	\frac{d \vec{\sigma_i}}{d t} = \mathbf{W}_i \vec{\sigma_i} + \sum_{j=1}^3 k_j [\vec{r}_j -(\vec{\sigma}_i \cdot \vec{r}_j) \vec{\sigma}_i]
	\label{eq3a}
\end{equation}
with
\begin{equation}
	\vec{r}_1 = \vec{r} = \frac{1}{N} \sum_{k=1}^N \vec{\sigma}_k,
	\label{vec1}
\end{equation}
\begin{equation}
	\vec{r}_2 =  \frac{2}{N} \sum_j (\vec{r} \cdot \vec{\sigma}_j) \vec{\sigma}_j  - \vec{r}
	\label{vec2}
\end{equation}	
and
\begin{equation}
	\vec{r}_3 = r^2 \vec{r}.
	\label{vec3}
\end{equation}
%

\section{Phase transition on the sphere and higher dimensions}

The transition to synchronization in $D=3$ is discontinuous in the case of all-to-all pair interactions, with $r=0$ if $k_1 < 0$ and $r(0^+) =1/2$ \cite{2019continuous}. The effects of $k_2$ and $k_3$ can be estimated with a 
mean-field calculation of the vectors $\vec{r}_2$ and $\vec{r}_3$.
Using Eq.~(\ref{vec2})  we write $\vec{r}_2 = \vec{\rho}_2 - \vec{r}$ with
\begin{equation}
	\vec{\rho}_2 = \frac{2}{N} \sum_j (\vec{r} \cdot \vec{\sigma}_j) \vec{\sigma}_j 
	                        =  \left(\frac{1}{N} \sum_j \vec{\sigma}_j \vec{\sigma}_j^T \right) 2 \vec{r}
	\label{sp5}
\end{equation}	
where the dyadic matrices are
\begin{equation}
	\vec{\sigma}_j \vec{\sigma}_j^T = \left(
	\begin{array}{ccc}
		\sigma_{jx}^2 & \sigma_{jx} \sigma_{jy} & \sigma_{jx} \sigma_{jz} \\
		\sigma_{jy} \sigma_{jx} & \sigma_{jy}^2 & \sigma_{jy} \sigma_{jz} \\
		\sigma_{jz} \sigma_{jx} & \sigma_{jz} \sigma_{jy} & \sigma_{jz}^2 
	\end{array}
	\right).
	\label{sp6}
\end{equation}
For asynchronous states we can assume that the components of the $\vec{\sigma}_j$ are uncorrelated and that the averages of $\sigma_{jx}^2$, $\sigma_{jy}^2$ and 
$\sigma_{jz}^2$ are $1/3$, since they sum to one. Thus, the average in Eq.~(\ref{sp5}) is $1/3$ times the identity matrix, and $\vec{\rho}_2 = 2 \vec{r} /3$ and $\vec{r}_2 = \vec{\rho}_2 - \vec{r} = - \vec{r}/3$.  Therefore, the net effect of the third order correction is to shift $k_1 \rightarrow k_{12} \equiv  k_1 - k_2/3$. The first order phase transition that would occur at $k_1=0$ is now shifted to the line $k_{12}=0$, or to
\begin{equation}
	k_1 = k_2/3.
	\label{sp7}
\end{equation}

To the right of this critical line the $\vec{\sigma}_j$'s start to correlate until they all point in the direction of $\vec{r}$. In this region we can approximate $\vec{\sigma}_j \approx \vec{r}$ and get
\begin{equation}
	\vec{\rho}_2 =   \left(\frac{1}{N} \sum_j \vec{\sigma}_j \vec{\sigma}_j^T \right) 2 \vec{r} \approx 2 r^2 \vec{r}
	\label{sp8}
\end{equation}	
so that $\vec{r}_2 = r^2 \vec{r}$.

Since $\vec{r}_3 = r^2 \vec{r}$, it does not contribute when $r=0$ and, therefore, does not affect the critical line $k_2=3k_1$. In the synchronous region it contributes positively, facilitating synchronization. Together with the third order term, that acts similarly, the effect of the higher order terms is to shift the curves 
$r=r(k_1)$ to the left by making $k_1 \rightarrow k_{1eff} = k_1 - r^2(k_2 + k_3)$ if $k_{1eff} > k_2/3$, making synchronization more effective (larger values of $r$) than the original model with pair interactions. If we call $r(k_1,k_2,k_3)$ the 
equilibrium value of the order parameter for fixed values of the coupling constants, then
\begin{equation}
	r(k_1,k_2,k_3) = \left\{
	\begin{array}{ll}
		0  &\qquad {\mbox {\textrm{if}}} \qquad k_1 < k_2/3, \\
		r(k_1+r^2(k_2+k_3),0,0) & \qquad {\mbox {\textrm{if}}} \qquad k_1 > k_2/3 .\\
	\end{array}
	\right.
\end{equation}
Conversely, all curves collapse onto $r(k_1,0,0)$ if shifted to the left according to
\begin{equation}
	r(k_1-r^2(k_2+k_3),k_2,k_3) = r(k_1,0,0) 
	\label{eqshift}
\end{equation}
in the region where $r \neq 0$.

Using similar arguments we see that the analogue of Eq.~(\ref{sp6}) in $D$ dimensions results in $(1/D) \mathbf{I}$ in the disordered region, where $\mathbf{I}$ is the $D \times D$ identity matrix. This implies that $\vec{\rho}_2 = (2/D) \vec{r}$ and 
$\vec{r}_2 = -[(D-2)/D] \vec{r}$. In $D=2$ there is no displacement in the critical point \cite{skardal2020higher}, but it shifts the transition in all other dimensions. The critical coupling goes from $k_{1c}$ to $k_{1c} + k_2 (D-2)/D$.

In odd dimensions we expect the transition to be discontinuous, with the critical line shifted. In even dimensions the transition remains continuous, but ever steeper as $k_2$ and $k_3$ increases.

\section{Numerical results}

In this section we show numerical simulations of Eqs.~(\ref{eq3a}) for $D=3$ and $D=4$. We consider $N=2000$ particles and choose the matrices of natural frequency $\mathbf{W}_i$ such that all entries are Gaussian distributed around $\omega=0$ with unit mean square deviation $\Delta=1$. (See \cite{Fariello2024a} for phase diagrams with pair interactions and different values of $\Delta$ and $\omega$.) Initial conditions are uniformly distributed over the corresponding $3D$ or $4D$ sphere and the equations are integrated up to $t=2000$. The order parameter $r$ is computed as the average over the last $500$ time units. 

\begin{figure}[p]
\centering
	\includegraphics[width=\textwidth]{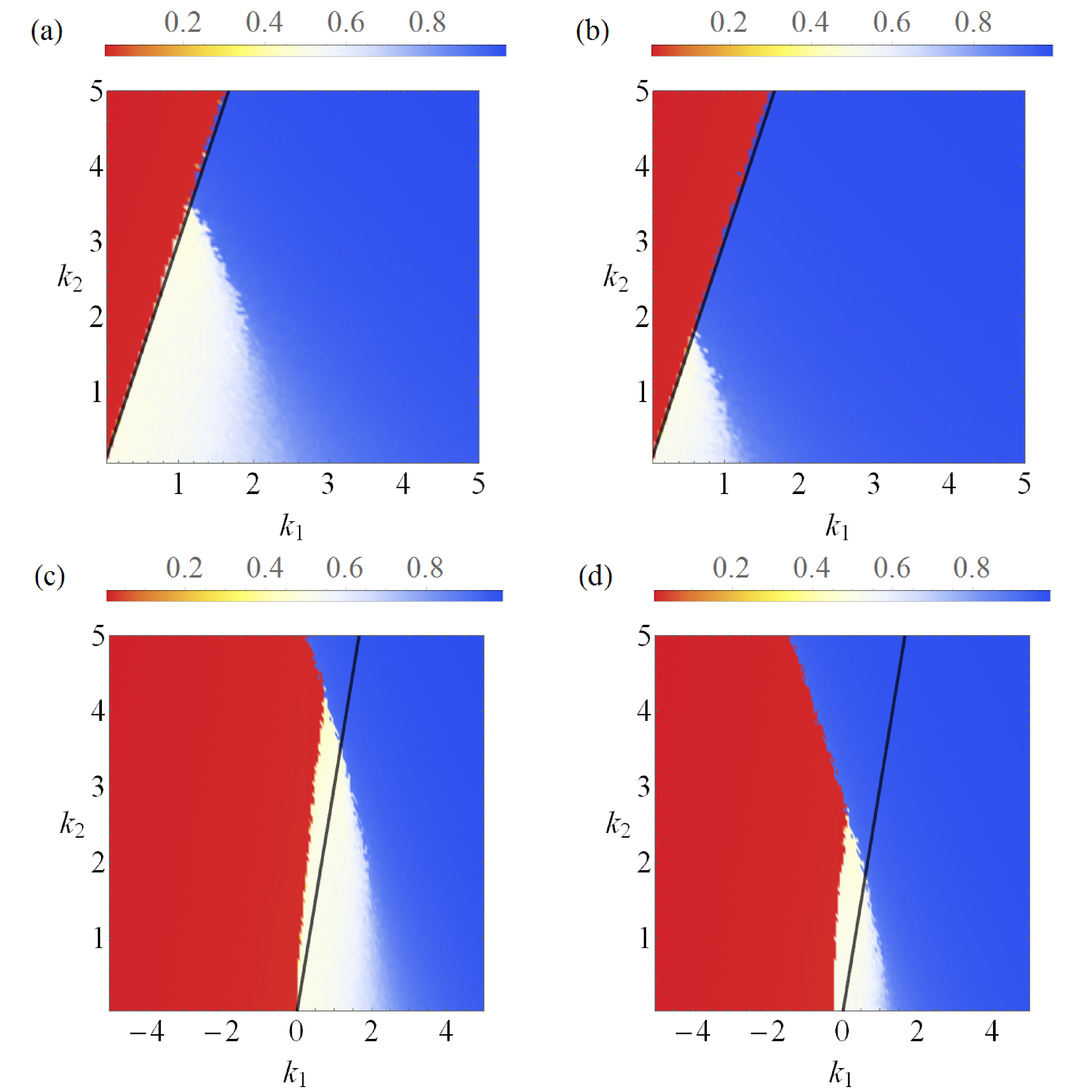}
	\caption{Heatmaps for $D=3$ showing the order parameter $r$ as a function of $k_1$ and $k_2$: (a) $k_3=0$; (b) $k_3=2$. In both cases oscillators have random initial conditions. The critical coupling for $k_2=0$ is $k_{1c}=0$, 
		where $r=1/2$, but it shifts to $k_{1c}=k_2/3$  (solid black line) when $k_2 \neq 0$. The value of $r$ right after the transition line increases as $k_2$ increases. Panels (c) and (d) show plots similar to (a) and (b), but starting with all oscillators at the same initial phase. Synchronous solutions exist even for $k_1<0$ if $k_2$ and $k_3$ are large enough. All values of $k_1$ and $k_2$ to the left of the solid black line where $r\neq 0$ correspond to the bi-stable regime.} 
	\label{fig1} 
\end{figure}

\begin{figure}[p]
\centering
	\includegraphics[width=\textwidth]{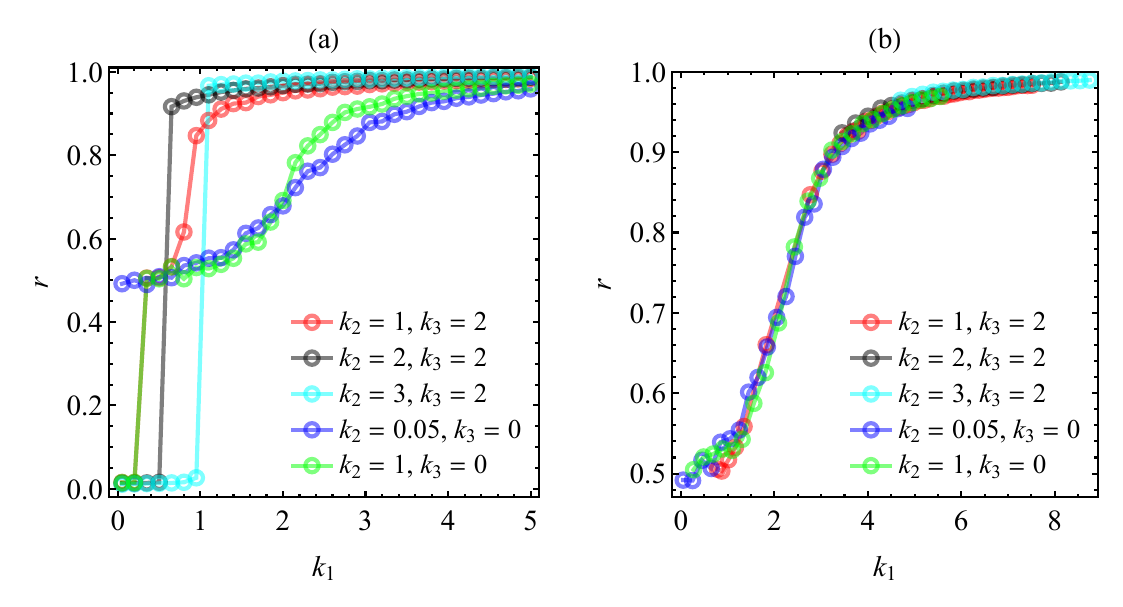}
	\caption{(a) Order parameter $r$ as a function of $k_1$ for different values of $k_2$ and $k_3$. The jump of size $1/2$ at $k_1=0$, when $k_2=k_3=0$, increases as $k_2$ and $k_3$ increase. (b) Curves for different values of $k_2$ and $k_3$ are shifted to the right according to Eq.~(\ref{eqshift}), collapsing on the curve 
		with $k_2=k_3=0$ when $r\neq 0$.} 
	\label{fig2}
\end{figure}

Figures~\ref{fig1} and \ref{fig2} show simulations for $D=3$. Figure~\ref{fig1}(a) shows heatmaps of the order parameter $r$ as a function of $k_1$ and $k_2$ for $k_3=0$ whereas Fig.~\ref{fig1}(b) shows results for $k_3=2$. In both panels the critical line from disorder to synchronization follows $k_2 = 3 k_1$ and is independent of $k_3$. The phase transition is always discontinuous, as in the case with $k_2=k_3=0$, but the
size of the discontinuity from $r=0$ to $r \neq 0$ increases from $1/2$ at $k_2=k_3=0$
to about $0.9$ when $k_2=4$ and $k_3=0$. For $k_3 > 0$ the jump increases even faster.

Figure~\ref{fig2}(a) shows $r$ as a function of $k_1$ for some values of $k_2$ and $k_3$. As $k_2$ increases the critical point moves towards larger values and the jump at the critical point increases. Figure~\ref{fig2}(b) shows the map 
$r(k_1,k_2,k_3) \rightarrow r(k_1-r^2(k_2+k_3),k_2,k_3)$, which shifts the curves to the left (when $r \neq 0$). All curves fall approximately on the same curve $r(k_1,0,0)$ as predicted by Eq.~(\ref{eqshift}).

\begin{figure}[p]
\centering
	\includegraphics[width=\textwidth]{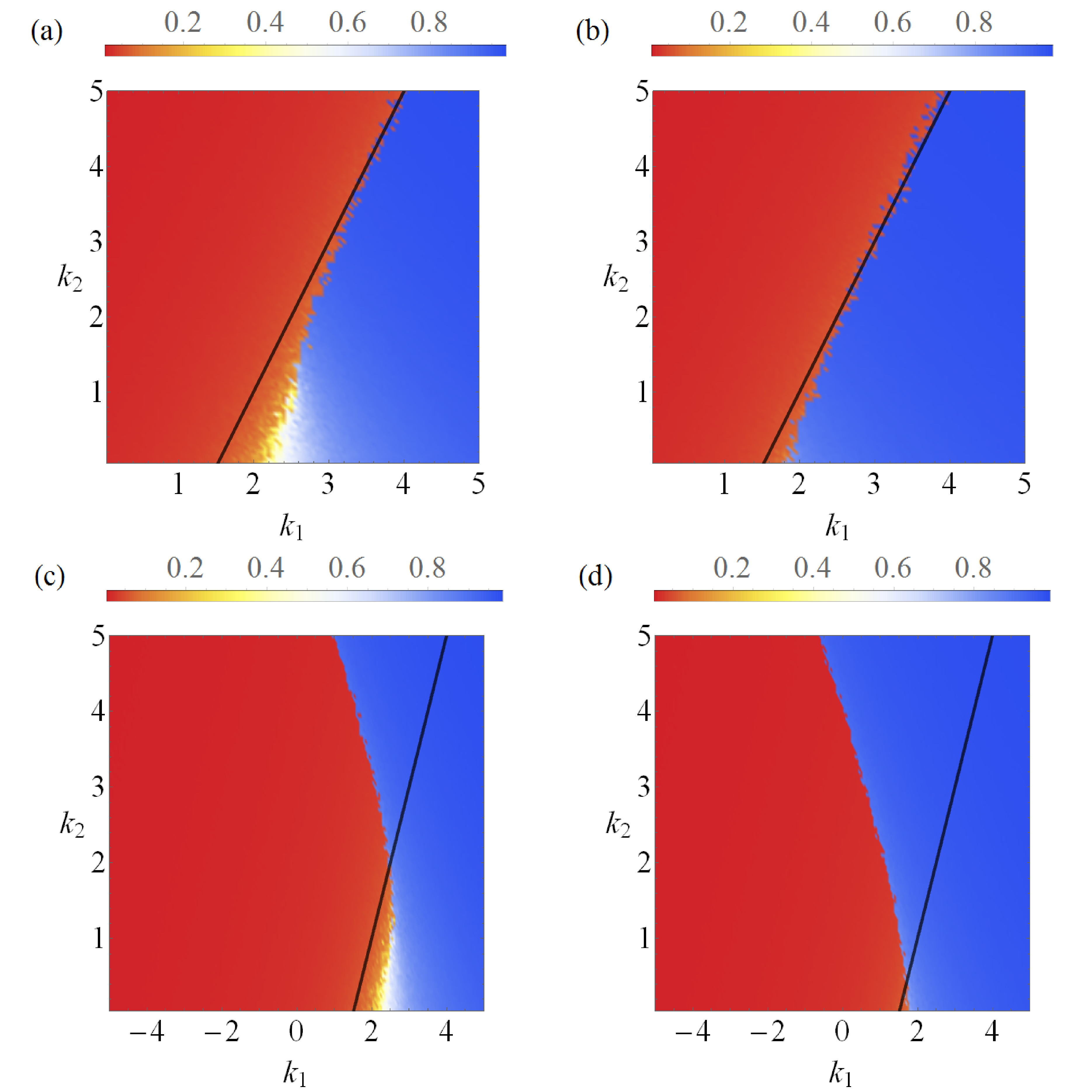}
	\caption{Heatmaps for $D=4$ showing the order parameter $r$ as a function of $k_1$ and $k_2$: (a) $k_3=0$; (b) $k_3=2$. In both cases oscillators have random initial conditions. The critical coupling for $k_2=0$ is $k_{1c} \approx 1.5$, but it shifts to $k_{1c}= 1.5+ k_2/2$ (solid black line) when $k_2 \neq 0$. 
		The value of $r$ right after the transition line increases as $k_2$ increases, leading to discontinuous transitions.
		Panels (c) and (d) show plots similar to (a) and (b), but starting with all oscillators at the same initial phase. Synchronous solutions exist even for $k_1<0$ if $k_2$ and $k_3$ are large enough. All values of $k_1$ and $k_2$ to the left of the solid black line where $r\neq 0$ correspond to the bi-stable regime.} 
	\label{fig3}
\end{figure}

\begin{figure}[p]
\centering
	\includegraphics[width=\textwidth]{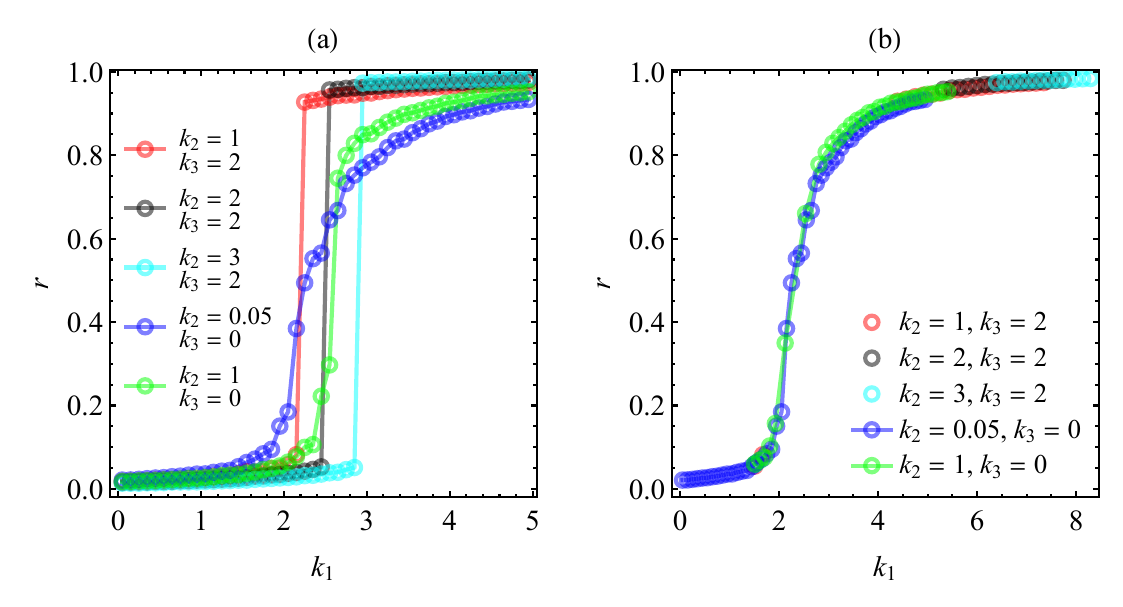}
	\caption{(a) Order parameter $r$ as a function of $k_1$ for different values of $k_2$ and $k_3$. The continuous transition at $k_{1c}=1.5$ when $k_2=k_3=0$ becomes discontinuous as $k_2$ and $k_3$ increase. (b) Curves for different values of $k_2$ and $k_3$ are shifted to the right according to Eq.~(\ref{eqshift}), collapsing on the curve with $k_2=k_3=0$ when $r\neq 0$.} 
	\label{fig4}
\end{figure}

Similar plots for $D=4$ are shown in Figs.~\ref{fig3} and \ref{fig4}. In this case the phase transition is continuous when $k_2=k_3=0$, with $k_{1c} \approx 1.5$. Again, the value of $k_3$ does not affect the shift in the critical coupling, which is
moved from $k_{1c}$ to approximately $k_{1c} + k_2/2$, as shown in Fig.~\ref{fig3}(a) for $k_3=0$ and Fig.~\ref{fig3}(b) for $k_3=2$. The approximation is still accurate, even in the neighborhood of the critical line. To the right of the critical line the system achieves higher values of synchronization as compared to the case $k_2=k_3=0$, leading to a discontinuity that increases as the intensity of higher order interactions increases, as shown in Fig.~\ref{fig4}(a). Figure~\ref{fig4}(b) shows that shifting the curves according to Eq.~(\ref{eqshift}) also makes them all collapse into $r(k_1,0,0)$.

\section{Bi-stability}

Higher order terms can give rise to hysteresis and bi-stability in $D=2$ \cite{skardal2020higher}. For example, for fully connected oscillators and Lorentzian distribution of natural frequencies, such that the Ott-Antonsen ansatz \cite{Ott2008}
can be employed, stable synchronized motion appears for sufficiently large values of $k_2+k_3$ even for $k_1 < k_{1c}$, where the incoherent state with $r=0$ is also stable.

In this section, we show, numerically, that this is also the case in dimensions $D=3$ and $D=4$. In order to obtain such synchronized states we generated the heatmaps shown in Figs.~\ref{fig1}(a), \ref{fig1}(b), \ref{fig3}(a) and \ref{fig3}(b) again, this time starting from all oscillators having the same initial phase. If a partially synchronized stable state exists, then such initial condition should be in its basin of attraction. 

Panels (c) and (d) of Fig.~\ref{fig1} show heatmaps of $r$ in the $k_1 \times k_2$ plane for $k_3=0$ and $k_3=2$ for $D=3$. The thick black line is the same as in panels (a) and (b), showing the transition from async to sync states starting close
to the basin of attraction of the $r=0$ solution. All states to the left of the black line with $r \neq 0$ co-exist with the $r=0$ solution and both are stable. The region of bi-stability increases with $k_2$ and $k_3$, similar to the 2$D$ case. Panels (c) and (d) of Fig.~\ref{fig3} show similar heatmaps for $D=4$. Again, regions of bi-stability develop for sufficiently large values of $k_2$ and $k_3$.

\section{Conclusions}

The effect of higher order interactions on the dynamics of complex systems has been a hot topic of research in the past years. Although this type of many body action depends in a great measure on the topology of connections between agents, mean-field estimates can serve as a guide to understand the changes expected when these interactions are included. Here we considered the multidimensional version of the Kuramoto model as introduced in \cite{2019continuous} and derived the functional form of third and fourth order corrections that are natural extensions of their counterparts in the usual, $D=2$, Kuramoto model.

We have shown that the third order correction has a special role in displacing the critical coupling constant towards higher values in all dimensions, except $D=2$. After the critical points, higher order terms tend to facilitate synchronization in
a simple way, shifting the order parameter $r(k_1,k_2,k_3)$ to $r(k_1+r^2(k_2+k_3),0,0)$. The shift causes the phase transition to become discontinuous also in even dimensions, where the original model
predicts continuous phase transitions.

Finally, we have shown that third and forth order interactions lead to bi-stability and hysteresis, similar to what was found in $D=2$, where synchronous solutions exist even for $k_1<0$.

\begin{acknowledgments}
	This work was partly supported by FAPESP, grant 2021/14335-0 (ICTP‐SAIFR) and CNPq, grant 301082/2019‐7. 
\end{acknowledgments}

\clearpage
\enlargethispage{1\baselineskip} 

\clearpage

\end{document}